\documentclass[aps,prd,showkeys,nofootinbib,superscriptaddress,preprint]{revtex4-1}
\usepackage{enumerate}
\usepackage{float}
\usepackage{amsmath,slashed,tensor,amssymb,epsfig,amsthm,bm,graphicx,graphics,slashed}
\usepackage[caption=false]{subfig}
\usepackage{hyperref}
\usepackage[utf8]{inputenc}
\usepackage[top=1.5cm, bottom=1.5cm, left=2cm, right=2cm]{geometry}

\usepackage{makecell}
\usepackage{microtype}
\usepackage[font=footnotesize,labelfont=bf]{caption}
\newcommand{\be}{\begin{equation}}
\newcommand{\ee}{\end{equation}}
\usepackage{color}
\definecolor{darkgreen}{rgb}{0,0.35,0}
\newcommand{\rc}{\textcolor{red}}

\newcommand{\ucharlesipnp}{Institute of Particle and Nuclear Physics, Charles University, V Hole\v{s}ovi\v{c}k\'ach 2, 18000 Prague 8, Czech Republic}
\newcommand{\ucharlestp}{Institute of Theoretical Physics, Charles University, V Hole\v{s}ovi\v{c}k\'ach 2, 18000 Prague 8, Czech Republic}
\newcommand{\eli}{Institute of Physics of the ASCR, ELI Beamlines Project, Na Slovance 2, 18221 Prague, Czech Republic}
\newcommand{\icfo}{ICFO-Institut de Ciences Fotoniques, The Barcelona Institute of Science and Technology, Avenue Carl Friedrich Gauss 3, 08860 Castelldefels (Barcelona), Spain}

\begin{document}
%-------------------------------------------------------------------------------------------------------------------

\title{Torsion in quantum field theory  \\ through time-loops on Dirac materials}
%\title{Time loops and torsion in Dirac materials}

\author{Marcelo F. Ciappina}
\email{marcelo.ciappina@eli-beams.eu}
\affiliation{\eli}
\affiliation{\icfo}
\author{Alfredo Iorio}
\email{iorio@ipnp.troja.mff.cuni.cz}
\affiliation{\ucharlesipnp}
\author{Pablo Pais}
\email{pais@ipnp.troja.mff.cuni.cz}
\affiliation{\eli}
\affiliation{\ucharlesipnp}
\author{Adamantia Zampeli}
\email{azampeli@utf.mff.cuni.cz}
\affiliation{\ucharlestp}

%\address{Faculty of Mathematics and Physics, Charles University \\
%V Hole\v{s}ovi\v{c}k\'ach 2, 18000 Prague - Czech Republic}

%\author{}
%
%\address{Faculty of Mathematics and Physics, Charles University \\
%V Hole\v{s}ovi\v{c}k\'ach 2, 18000 Prague - Czech Republic}
%\\
%and \\
%Department of Physics, KU Leuven Campus Kortrijk – Kulak \\
%Etienne Sabbelaan 53 bus 7657, 8500 Kortrijk, Belgium}
%
%\author{}
%
%\address{Faculty of Mathematics and Physics, Charles University \\
%V Hole\v{s}ovi\v{c}k\'ach 2, 18000 Prague - Czech Republic}
%-------------------------------------------------------------------------------------------------------------------
%-------------------------------------------------------------------------------------------------------------------
\begin{abstract}
Assuming dislocations could be meaningfully described by torsion, we propose here a scenario based on the role of time in the low-energy regime of two-dimensional Dirac materials, for which coupling of the fully antisymmetric component of the torsion with the emergent spinor is not necessarily zero. Appropriate inclusion of time is our proposal to overcome well-known geometrical obstructions to such a program, that stopped further research of this kind. In particular, our approach is based on the realization of an exotic \textit{time-loop}, that could be seen as oscillating particle-hole pairs. Although this is a theoretical paper, we moved the first steps toward testing the realization of these scenarios, by envisaging \textit{Gedankenexperiments} on the interplay between an external electromagnetic field (to excite the pair particle-hole and realize the time-loops), and a suitable distribution of dislocations described as torsion (responsible for the measurable holonomy in the time-loop, hence a current). Our general analysis here establishes that we need to move to a nonlinear response regime. We then conclude by pointing to recent results from the interaction laser-graphene that could be used to look for manifestations of the torsion-induced holonomy of the time-loop, e.g., as specific patterns of suppression/generation of higher harmonics.
\end{abstract}

%\pacs{***}


\maketitle

\section{Introduction}
\label{section_introduction}

To date, there is no experimental evidence of torsion of spacetime and the most prominent theory of gravity we have, Einstein's General Theory of Relativity, does not contemplate torsion. Nonetheless, torsion remains the focus of important research, both in fundamental and in condensed matter physics.

On the fundamental side, just like curvature is intimately connected with mass, torsion is intimately connected with spin, see, e.g., the pioneering work of Kibble \cite{KibbleTorsion1961}. Some argue that torsion manifests itself through the very existence of spinors, in an otherwise standard spacetime (see, e.g., \cite{CartanTheoryHehl}), while others continue to pursue the idea that torsion may as well be an actual physical property of our spacetime, within an extended theory of gravity (see, e.g., \cite{Hehl:1976kj}) or of quantum gravity (see, e.g, \cite{PercacciTorsion}). Furthermore, both standard Supersymmetry (SUSY), in its curved space
declination (supergravity, SUGRA) \cite{WessBagger} and the more recent unconventional SUSY  (USUSY) \cite{AVZ} make extensive use of torsion.

On the condensed matter side, the existence of two kinds of basic topological defects, disclinations and dislocations, related respectively to curvature and torsion, makes it natural to include torsion in the geometrical description of the physical properties of materials \cite{Katanaev:1992kh,Kleinert_book}. This is not entirely free of ambiguities, in particular when it comes to associate a specific torsion to a given distribution of Burgers vector; but surely torsion is one of the two geometric entities at work there, along with curvature.

In the last years, due to their low energy spectrum structure, Dirac materials \cite{WehlingDiracMaterials2014} have emerged as experimental playgrounds where both kinds of arenas, the fundamental research and the condensed matter one, met. In particular, the role of disclinations is under intense investigation to realize graphene analogs of Dirac quantum fields in curved spacetimes, see, e.g., \cite{IORIO20111334,iorioWeylLab2012, Iorio:2011yz,Iorio:2013ifa,Iorio:2015iha,Iorio:2017vtw,IorioPais} and recently the role of yet another kind of defects (grain boundaries) was also explored \cite{IORIO2018265}. Investigations on how, in this context, dislocations could be used to construct an analog Dirac field theory coupled with torsion, rather than curvature, were of course carried on, see, e.g., \cite{Mesaros:2009az}.

If we were able to do so, it would be an invaluable help to shed light on some of the above recalled mysteries on torsion. Let us mention, for instance, USUSY, especially in its $(2+1)-$dimensional formulation, that has been found to have many similarities with the Dirac field theory on graphene, see \cite{SU2Ususy,GPZ}, and especially the recent \cite{DauriaZanelli2019}. Unfortunately, the exploration of the role of torsion in this setting found a geometric obstacle, just due to the $2+1$ dimensions: As we shall recall later, a Dirac spinor only couples to the fully antisymmetric component of torsion, hence three dimensions are necessary. Lacking the spatial third dimension, this seemed impossible \cite{deJuan2010,Vozmediano:2010zz,Amorim:2015bga}. These ``no-go'' results stopped research in this direction. It is the main goal of this work to suggest a way to surmount this obstacle, based on the use of time as the necessary third dimension.

In what follows we shall first recall, in Section \ref{section_torsion}, how the geometrical obstruction to have torsion in two-dimensional Dirac materials comes about, while in Section \ref{section_t_loop} we propose our way to overcome it. In Section \ref{section_macro_extraction_from_micro} we explore the possible ways to extract experimental data from the condensed matter system related to the microscopic analog relativistic model. In Section \ref{section_linear} we put some flesh on the latter bones, by identifying the responses to combined electromagnetic and disclination/torsion perturbations. In Section \ref{section_nonlinear} we point to the nonlinear response regime as the one necessary to find the effect we are looking for. Finally, in the concluding Section \ref{section_conclusions}, we summarize the results and point to future work.

\section{Torsion and two-dimensional Dirac materials}
\label{section_torsion}

By definition, Dirac materials's $\pi$ electrons\footnote{In the following we refer to two dimensional Dirac materials, with hexagonal lattice. Examples are graphene, germanene, silicene \cite{WehlingDiracMaterials2014}.} obey a low-energy dynamics near a Dirac point, governed by an emergent relativistic-like Hamiltonian with structure $H_D = v_F \vec{\sigma} \cdot \vec{p}$, where $v_F$ is the Fermi velocity, and vectors are spatial two dimensional, see, e.g., \cite{PacoReview2009}. To fully take into account this emergent relativistic-like structure \cite{IORIO20111334}, we include time as $x^0 = v_F t$, hence turn to the $(2+1)$-dimensional action\footnote{We use Latin indices $a,b,\ldots$ for tangent/flat space and Greek indices $\mu,\nu,\ldots$ for base curved manifold. We choose the signature $\eta_{ab}=\mbox{diag}(+,-,-)$. The vielbeins are denoted by $e^{a}_{\mu}$ and their inverse by $E_{a}^{\mu}$.}
\begin{equation}\label{flatAction}
S_0 [\overline{\Psi}, \Psi] = i \hbar v_F \int d^{3}x  \overline{\Psi}\gamma^{a}\partial_{a}\Psi \,.
\end{equation}
Here, the Dirac spinor is not in the irreducible representation of the Lorentz group $SO(1,2)$, it has four components $\Psi = (\psi_+, \psi_-)^T$, with $\psi_\pm = (\alpha_\pm, \beta_\pm)^T$. The variables $\alpha$ and $\beta$ denote the sublattice anticommuting operators, acting near the two inequivalent Dirac points labelled with ``$\pm$''. This might seem a redundancy, especially for the choice of the Clifford algebra presented in Appendix \ref{appendix_torsion}, for which the two two-component spinors are fully decoupled. Nonetheless, this is the general setting one must use because the interaction we are about to consider might, in principle, as well couple the two irreducible spinors. This will be discussed later here, and more details can be found in \cite{IORIO2018265} on the role of the two Dirac points, and on the various choices for the Clifford algebra.

\begin{figure}
\begin{center}
\subfloat[][Screw Dislocation]{\includegraphics[width=.4\textwidth,angle=90]{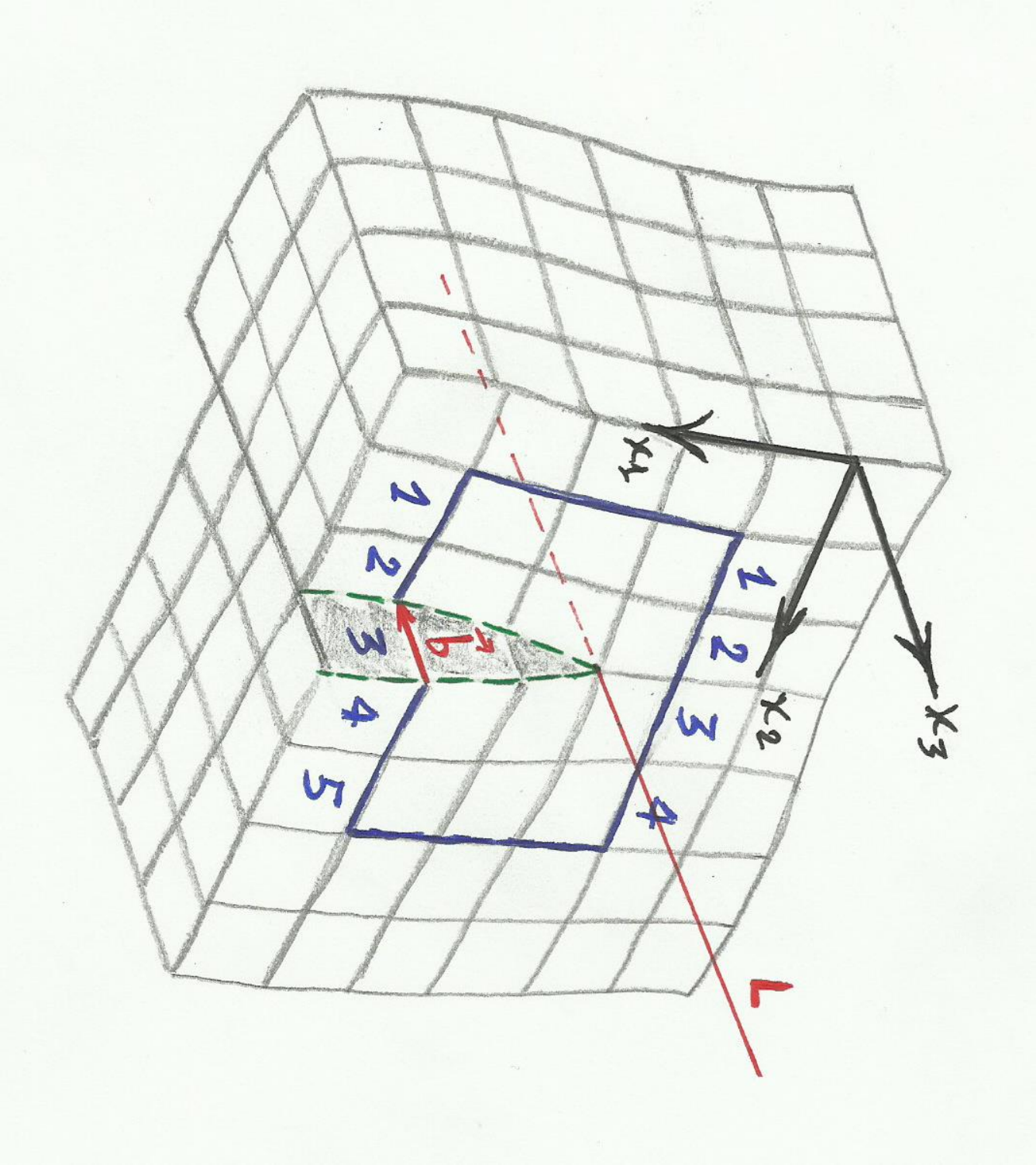}}
\subfloat[][Edge Dislocation]{\includegraphics[width=.4\textwidth,angle=90]{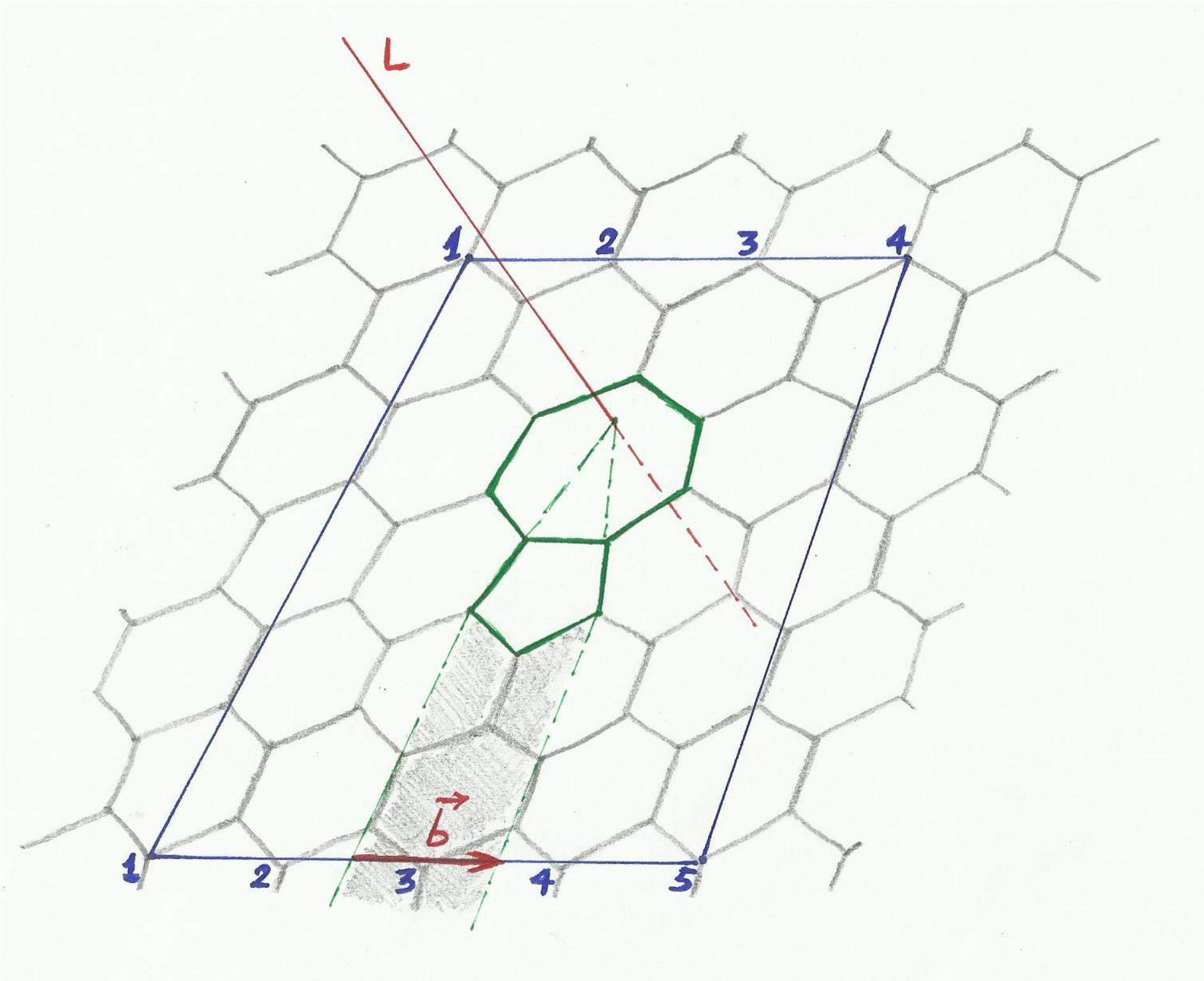}}
\caption{ (i) \emph{Screw dislocation} in a cubic lattice. Burgers vector and the dislocation line are parallel here. The circuit presents an extra step when $\vec{b}$ is nonzero. For $\chi_3 \equiv t$ this configuration could give rise to nonzero temporal components of torsion, an instance to be investigated in the context of the ``time crystals'' of \cite{Wilczek2012}. (ii) \emph{Edge dislocation} in an hexagonal two-dimensional lattice, typical of a vast class of Dirac materials \cite{WehlingDiracMaterials2014}. The Burgers vector, $\vec{b}$, lies in the plane, while the dislocation line, $L$, is perpendicular to it, hence always orthogonal to $\vec{b}$. For this particular case of a pentagon-heptagon pair, this configuration is also called \emph{glide dislocation} \cite{deJuan2010}. To close the circuit, with this $\vec{b} = (1,0)$, the number of steps (five here) is larger by one unit for the portion that includes the shaded area, with respect to portion running in the defect-free part (four steps here).}\label{Fig1}
\end{center}
\end{figure}

The natural generalization of (\ref{flatAction}) to a $(2+1)$-dimensional spacetime, equipped with a metric $g_{\mu\nu}=\eta_{ab}e^{a}_{\mu}e^{b}_{\nu}$ and a metric-compatible connection $\Gamma^{\lambda}_{\mu \nu}$ \textit{that includes torsion} \cite{Nakahara}
\begin{equation}
T^{\lambda}_{\mu \nu} = \Gamma^{\lambda}_{\mu \nu} - \Gamma^{\lambda}_{\nu \mu} \,,
\end{equation}
is
\begin{equation}
S = i \hbar v_F  \int d^{3}x \sqrt{-g} \overline{\Psi}\gamma^{\mu} D_{\mu}\Psi,
\end{equation}
where the covariant derivative is defined as $D_{\mu}\Psi=\partial_{\mu}\Psi+\frac{i}{2}\omega^{ab}_{\mu}\mathbb{J}_{ab}\Psi$, with $\mathbb{J}_{ab}=\frac{i}{4}[\gamma_{a},\gamma_{b}]$ the Lorentz generators in spinor space. The spin-connection, $\omega^{ab}_{\mu}=e^{a}_{\lambda}(\delta^{\lambda}_{\nu}\partial_{\mu}+\Gamma^{\lambda}_{\mu\nu})e^{b\nu}$, can be decomposed into torsion-free and {\it contorsion} contributions \cite{Z-book}, $\omega_\mu^{ab}=\mathring{\omega}_\mu^{ab}+ \kappa_\mu^{ab}$, where $T^{\lambda}_{\mu\nu}=E_a^{\lambda}\tensor{\kappa}{_{\nu}^{a}_{b}}e^{b}_{\mu}-E_a^{\lambda}\tensor{\kappa}{_{\mu}^{a}_{b}}e^{b}_{\nu}$. Standard manipulations of the action $S$, reported in detail in the Appendix \ref{appendix_torsion}, lead to the form
\begin{equation}\label{action_torsion}
S = i \hbar v_F\int d^{3}x \; |e| \; \overline{\Psi}\left(\gamma^{\mu}\mathring{D}_{\mu} - \frac{i}{4} \gamma^{5} \frac{\epsilon^{\mu\nu\rho}}{|e|} T_{\mu \nu\rho} \right)\Psi\;,
\end{equation}
where $|e|=\sqrt{|g|}$, the covariant derivative, $\mathring{D}_{\mu}$, is based on the torsion-free connection, $\mathring{\omega}_\mu^{ab}$, only, $\gamma^{5} \equiv  i \gamma^{0} \gamma^{1} \gamma^{2} = \left(
                                                           \begin{array}{cc}
                                                             I_{2 \times 2} & 0 \\
                                                             0 & -I_{2 \times 2} \\
                                                           \end{array}
                                                         \right)$
(we used the conventions of \cite{IORIO2018265} for $\gamma^{0},\gamma^{1},\gamma^{2}$ giving a $\gamma^{5}$ that {\it commutes} with the other three gamma matrices\footnote{This is due to the reducible, rather than irreducible, representation of the Lorentz group we use.}), and the contribution due to the torsion is all in the last term through its totally antisymmetric component \cite{Shapiro}. From here, it is evident that the emergent fermions of Dirac materials $\Psi$ can only be coupled to the component $T_{012}$ (or also with $T_{102}$, or $T_{210}$). This is the above-mentioned geometric obstacle, that led earlier investigators to conclude that, for two dimensional Dirac materials, dislocations could not be accounted for by torsion \cite{deJuan2010,Vozmediano:2010zz,Amorim:2015bga}.

Notice that, when odd-sided defects of the kind indicated in the right picture of Fig.1 are present, the two triangular sublattices, say $A$ and $B$, making the hexagonal lattice of graphene (and of other Dirac materials of the family), get intertwined at specific locations, where a sort of frustration occurs (that is, those points belong to both A and B at the same time). As noticed in \cite{IORIO2018265} (see also \cite{GonzalezFullerenePRL}), for some particular descriptions, the two Dirac points are related by a change in the sublattice, $A \leftrightarrow B$. While this only happens for certain suitable descriptions (e.g., even for the same Dirac points we have different Hamiltonian choices, see Appendix B of \cite{IORIO2018265}), since the actual physics is independent from this choice, we can always pick up a description where this is true. Therefore, in general, when such defects are present we must use both Dirac points, as we do here.

The torsion tensor in crystals is related to the Burgers vector through the formula\footnote{Despite the apparent simplicity of the formula \eqref{torsion-Burgers}, in practice it is a difficult task to assign the torsion tensor for a given distribution of Burgers vector on the graphene sheet, see, e.g., \cite{Lazar2003,banhart2010structural}.} \cite{Kleinert_book,Katanaev2005}
\begin{equation}\label{torsion-Burgers}
b^{a}=\int\int_{\Sigma} e^{a}_{\lambda}T^{\lambda}_{\mu \nu}dx^{\mu} \wedge dx^{\nu} \;,
\end{equation}
where $\Sigma$ is a surface containing the dislocation, but otherwise arbitrary, $a = 0,1,2$. We clearly see that the only two possibilities that a nonzero Burgers vector can give rise to $\epsilon^{\mu\nu\rho}T_{\mu \nu\rho}\neq0$, necessary for the
coupling in (\ref{action_torsion}), are (cf. Fig.~\ref{Fig1}): (i) a \textit{time directed} screw dislocation, i.e. $b_{t} \propto \int\int  T_{012} dx \wedge dy$ or (ii) an edge dislocation spotted by a \textit{space-time circuit}, e.g, $b_{x} \propto \int\int  T_{102} dt \wedge dy$. Here we took $e^{a}_{\mu}=\delta^{a}_{\mu}$, in both circumstances.

\section{Using ``time loops'' to overcome the two-dimensional geometric obstruction}
\label{section_t_loop}

Our claim here is that both scenarios, are in fact not impossible. The first scenario could be explored in the context of the fascinating time crystals introduced by Wilczek \cite{Wilczek2012, WilczekShapere2012}, and it is the focus of intense experimental studies (see, e.g., \cite{PhysRevLett.109.163001} and the recent \cite{PhysRevLett.121.185301}). Such lattices, discrete in all dimensions, including time, would be an interesting playground
to probe ideas of quantum gravity \cite{Loll1998}, although in $2+1$ dimensions\footnote{In $2+1$ dimensions we do have a defect-based approach to classical gravity/geometry, see \cite{Katanaev:1992kh} and \cite{Kleinert_book}, although the role of time is not clear there.}. In what follows, we shall not focus on this, but rather on the second scenario.

In the Appendix \ref{appendix_zero_curvature} we show that we can take the Riemann curvature to be zero, $\mathring{R}_{\mu \nu}^{ab}=0$, but with $\kappa_\mu^{ab}\neq0$, and choose a frame where $\mathring{\omega}_\mu^{ab}=0$. These settings make possible to isolate the effects of torsion on the system, and the corresponding action is
\begin{equation}\label{action_pure_torsion}
S = i\hbar v_F \int d^{3}x |e| \,  \left(\overline{\Psi}\gamma^{\mu}\partial_{\mu}\Psi - \frac{i}{4} \overline{\psi}_{+} \phi \psi_{+} + \frac{i}{4} \overline{\psi}_{-} \phi \psi_{-} \right) \;,
\end{equation}
where $\phi \equiv \frac{\epsilon^{\mu \nu \rho}}{|e|}T_{\mu\nu\rho}$. As clearly shown in \eqref{action_pure_torsion}, even in the presence of torsion, the two irreducible spinors, $\psi_{+}$ and $\psi_{-}$, actually decoupled. Nonetheless, they couple to $\phi$ with opposite signs.

\begin{figure}
\begin{center}
\includegraphics[width=0.4\textwidth,angle=90]{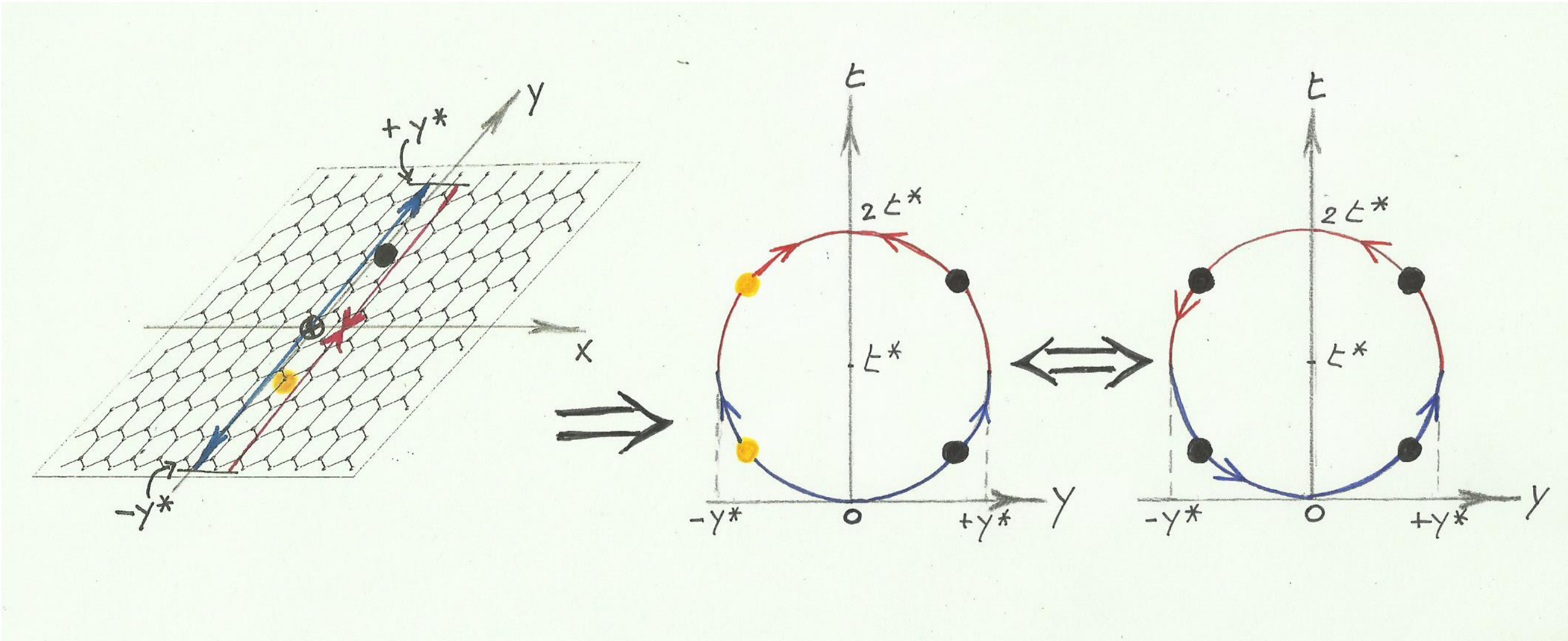}
\end{center}
\caption{Idealized \textit{time-loop}. At $t=0$, the hole (yellow) and the particle (black) start their journey from $y=0$, in opposite directions. Evolving forward in time, at $t=t^*>0$, the hole reaches $-y^*$, while
the particle reaches $+y^*$,  (blue portion of the circuit). Then they come back to the original position, $y=0$, at $t=2t^*$ (red portion of the circuit). This can be repeated indefinitely. On the far right, the
equivalent \textit{time-loop}, where the hole moving forward in time is replaced by a particle moving backward in time.}%
\label{Fig3TIMELOOP}%
\end{figure}

To spot the effects of $\phi$, we propose to make use of the particle-antiparticle structure, encoded in the action (\ref{action_pure_torsion}). Indeed, the regime of Dirac materials we describe, is the ``half-filling'' \cite{PacoReview2009}, whose vacuum state has the vacancies of the valence band ($E < 0$) completely filled, and the vacancies of the conduction band ($E > 0$) empty. This is the analog of the Dirac sea. If a pair particle-hole is excited out of this vacuum, and particle and antiparticle are made to oscillate, say, along $y$, as described in Fig.~\ref{Fig3TIMELOOP}, this amounts to a circuit of the particle-antiparticle pair in the $(y,t)$-plane. Fully exploiting the emergent relativistic-like structure of the model, the portion of the circuit described by the \textit{\textit{antiparticle}} moving \textit{forward} in time, corresponds to the \textit{particle} moving \textit{backward} in time. This realizes a \textit{time-loop}. The pictures in Fig.~\ref{Fig3TIMELOOP} refer to a defect-free sheet. The presence of a dislocation, e.g., like the one in Fig.~\ref{Fig1}, with Burgers vector $\vec{b}$ directed along $x$, would result in a failure to close the loop proportional to $\vec{b}$.

Within the idealized, single-particle/classical picture, what we are saying is that: provided dislocations can be meaningfully described by a suitable torsion tensor, the low energy Dirac field theory emerging here can include a nonzero coupling with torsion, accounting for a field theory description of the effects of dislocations, only when the third dimension is taken to be time. This is a nice idea, but the real challenge is to bring this idealized picture close to experiments. We introduce below the first steps in that direction.

\section{Bridge between the microscopic/classical picture and its macroscopic measurable manifestations}
\label{section_macro_extraction_from_micro}

The primary aim of this paper is theoretical. Namely, as just recalled, we wanted to point to a way to overcome a geometric obstruction, via the inclusion of time in the picture. Nonetheless, we shall now move some steps toward clarifying how to extract measurable effects of this microscopic picture. In other words, we shall illustrate the steps involved in going from the microscopic classical one-particle action we propose, till, e.g., an ammeter measuring a macroscopic current  that is the manifestation of the effect.

Generally speaking, there are four families of experiments one could perform on our Dirac material, based on the following class of phenomena: i) thermodynamics; ii) spectroscopy; iii) thermal and electronic transport and iv) Scanning Tunneling Microscopy (STM). In this Section we shall try to be as general as possible, and keep in mind all possibilities, although the first family of phenomena is perhaps the less suitable, because, when considering thermodynamics, the microscopic properties are, so to speak ``averaged away''. On the other hand, from the list above, the experiments on the electronic transport seem the most appropriate, because the quasi-particles we are describing are indeed those responsible for such properties. Hence, at the end of this road, we shall indeed be seeking for experiments on the transport properties.

Nonetheless, we want to keep the generality as much possible here, for two reasons. First, since one might envisage different roads than those we have in mind, we want to furnish the first aid there too. Second, we want to clarify one important aspect of our approach, that is the interplay between different languages, typically at work when dealing with analogs. On the one hand, we shall have the field theoretical description of relativistic systems, on the other hand we shall have the condensed matter description. We shall declare which is the language in use, case by case, as clearly as possible, although sometimes we may forget, or deem it to be self-evident. Hence, this warning, at this stage of the paper, should alert the reader to pay the due attention to this delicate point, from here on.

In our view, the simplest settings to realize in practise the microscopic picture above presented, need: i) an {\it external electromagnetic field} to excite the pair particle-hole necessary for the time-loop, and ii) that a suitable disclination/torsion provides the non-closure of the loop in the appropriate direction, something we shall refer to as {\it holonomy}. To summarize: we are looking for \textit{the measurable effects of a disclination/torsion-induced holonomy in a time loop}. It is only a (suitable) \textit{combination} of those interactions that can produce the effect we are looking for.

Therefore, the action governing the relevant microscopic dynamics is
\begin{eqnarray}
S  & = & i  \int d^{3}x \,|e|\, \left(\overline{\Psi} \gamma^{\mu} (\partial_{\mu} - i g_{\mbox{em}} A_\mu) \Psi - i g_{\mbox{tor}} \overline{\psi}_{+} \phi \psi_{+} + i g_{\mbox{tor}} \overline{\psi}_{-} \phi \psi_{-} \right) \;, \label{action torsion external A} \\
& \to & i  \int d^{3}x \, \left(\overline{\psi} \gamma^{\mu} \partial_{\mu} \psi  - i g_{\mbox{em}} \hat{j}^{\mu}_{\mbox{em}} A_\mu - i g_{\mbox{tor}} \hat{j}_{\mbox{tor}} \phi \right) \equiv S_0 [\overline{\psi}, \psi] + S_I [A, \phi] \;. \label{action samples}
\end{eqnarray}
where, we have set constants to one, $g_{\mbox{em}}$ and $g_{\mbox{tor}}$ are the electromagnetic and torsion coupling constants, respectively. In the last line, to avoid unnecessary complications, we considered only one Dirac point, say $\psi \equiv \psi_+$, and the metric is taken to be flat, $|e|=1$, hence the indices are the flat ones, $\mu, \nu, ... \to a,b,...$, but nonetheless, to ease the notation, we shall use Greek letters, anyway. Finally, $\hat{j}^{\mu}_{\mbox{em}} \equiv  \overline{\psi} \gamma^{\mu} \psi$, while $\hat{j}_{\mbox{tor}} \equiv \overline{\psi} \psi$.

The electromagnetic field is \textit{external}, hence a four-vector\footnote{A different, if not more naturally $(2+1)-$dimensional setting would be to obtain $A_\mu$ by suitably straining the material, see, e.g., \cite{Vozmediano:2010zz,Amorim:2015bga}, and \cite{IorioPais}. In that case, a typical setting is $A_t \equiv 0$, $A_x \sim u_{xx} - u_{yy}$, $A_x \sim 2 u_{xy}$, where $u_{i j}$ is the strain tensor.} $A_\mu \equiv (V, A_x, A_y,A_z)$. Nonetheless, the dynamics it induces on the electrons living on the membrane is two-dimensional, therefore, the effective vector potential may be taken to be\footnote{Alternatively, the so-called {\it reduced QED} approach can be taken. In such approach, the gauge field propagates in a three-dimensional space and one direction is integrated out to obtain an effective interaction with the electrons constrained to move in a two-dimensional plane \cite{Marino:1992xi,Gorbar:2001qt}. This approach could shed some light on the appearance of a photon Chern-Simons term \cite{Dudal:2018mms,Dudal:2018pta}.} $A_\mu \equiv (V, A_x, A_y)$, see, e.g., \cite{PhysRevLett.121.207401, natureLaserGraphene}.
Similarly, the torsion field $\phi$ as well enters into the action as an external field, because there is no dynamical kinetic term for it, although there are no issues about dimensionality here. A different view, when $\phi$ is constant, is to include it into the unperturbed action, where it plays the role of a mass $S_0 \to S_m$, see, e.g., \cite{DauriaZanelli2019}, where $S_m = i \int d^{3}x  \; \overline{\psi}(\slashed{\partial} - m(\phi)) \psi$.

\begin{figure}
\begin{center}
\includegraphics[width=0.5\textwidth,angle=0]{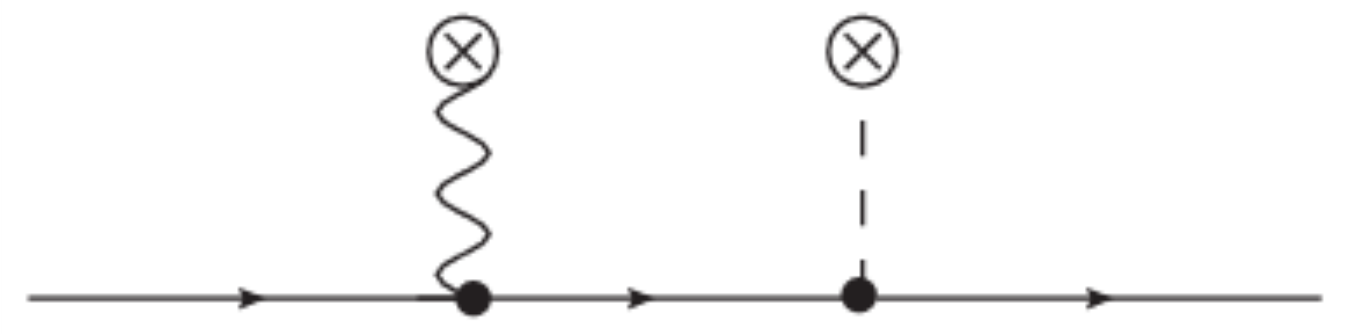}
\end{center}
\caption{Interaction of the $\pi$ electron with an external electromagnetic field (wavy line) and the external $\phi$ field (dashed line). This sketched and not literal diagram is the simplest possible nonzero contribution to the process of torsion-induced holonomy in the time-loop we are seeking.}%
\label{Fig_interaction}%
\end{figure}

With this in mind, the generic one-particle diagram that represents the microscopic phenomenon we are seeking, without taking it too literally as a real Feynman diagram, is given in Fig. \ref{Fig_interaction}. Of course, the details of the specific settings that give rise to the wanted torsion-induced holonomy in the time-loop, are all to be found. In what follows we shall establish constraints and general properties of these terms.

We are in the situation described by the microscopic perturbation
\begin{equation}\label{SIfx}
S_I [F_i] = \int d^{3}x \, \hat{X}_i (\vec{x},t) F_i (\vec{x},t) \;,
\end{equation}
with the system responding through $\hat{X}_i (\vec{x},t)$ to the external probes $F_i (\vec{x},t)$. The general goal is then to find
\begin{equation}\label{X[F]}
\hat{X}_i [F_i] \;,
\end{equation}
to the extent of predicting a measurable effect of the combined action of the two perturbations $F_i (\vec{x},t)$: $F^{\mbox{em}}_{1} (\vec{x},t) \propto A_\mu (\vec{x},t)$, that induces the response  $\hat{j}^\mu_{\mbox{em}}$, and $F^{\mbox{tor}}_{2} (\vec{x},t) \propto \phi (\vec{x},t)$, that induces the response  $\hat{j}_{\mbox{tor}}$:
\begin{equation}\label{SIaphi}
S_I [A, \phi] = \int d^{3}x \, \left( \hat{j}^\mu_{\mbox{em}} A_\mu + \hat{j}_{\mbox{tor}} \phi \right) \,,
\end{equation}
where we have included the couplings, $g_{\mbox{em}}$ and $g_{\mbox{tor}}$, in the respective currents.

The first comment we make is that we shall keep open the possibility of a time dependence for both perturbations, not only for the obvious electromagnetic one. Of course, the typical time scales involved are different, $\tau_{\mbox{tor}} \gg \tau_{\mbox{em}}$, but indeed the defects do have a dynamics. They can form, dissolve, move, as shown in various annealing processes. Nonetheless, we shall not consider that dynamics here, just like we do not include the dynamics of the electromagnetic field in our study of the response of the system.

A second general comment, is that, apart from $\gamma^0$, and the actual values of the couplings, the response to the voltage $V$ is of a similar type as the response to the $\phi$ perturbation, $\hat{j}^{0}_{\mbox{em}} \sim \hat{j}_{\mbox{tor}}$. As we shall recall later, this is a change in density of $\pi$ electrons, rather than a flowing of its current.

The third comment is related to the formal mathematical language used. We shall use the relativistic language of the actions above, and shall try to extract from it all possible predictions. Therefore, in the averaging processes, necessary to compare the predictions of the microscopic theory with the experiments, we shall use the functional integral methods, as they are the most natural in the relativistic context. Nonetheless, it is instructive and important to perform similar calculations in the density matrix/Hamiltonian approach, that is the most suitable for the tight-binding description. With the latter in hands, one would have an independent check of the relativistic predictions.

A fourth general comment is that the visible macroscopic effects we are seeking, should come from some form of mismatch between retarded and advanced Green functions, $G^R$, and $G^A$, respectively, referring to the propagations of particles, and holes, respectively. The relativistic picture, as known, uses the causal or Feynman propagator, that employs bits of both types of Green functions. These, along with the expectedly non trivial topological properties of the $\phi$ term, are delicate issues that need to be addressed to describe in mathematical terms the t-loop, and its torsion-induced holonomy in the proper language.

The last comment of this Section is a very important one, namely that the unperturbed system, described by $S_0$ (or, by $S_m$), is a free, non interacting system, hence, in principle, exactly solvable. Therefore, we see that the most delicate issues here are the appropriate boundary conditions. It is there that the details of the actual realization of the scenario of interest will emerge.

\section{The linear response regime}
\label{section_linear}

For the sake of this general discussion, and in order to learn the structure of the quantities involved, we shall now focus on weak perturbations. Hence, we can use the linear response which gives for (\ref{X[F]})
\begin{equation}\label{xlinear}
  X_i (\vec{x},t) = \int d^3 x' \chi_{i j} (\vec{x},t ; \vec{x}',t') F_j (\vec{x}',t') \; + \; O(F^2) \;,
\end{equation}
where $\chi_{i j}$ is the response function, which encodes the microscopic details of the system.

The macroscopic response, $X$, should be seen as the expectation value of some one-particle operator, both from the quantum and the statistical average point of view
\begin{equation}\label{doubleaverage}
  X = \sum_{a b} \langle \overline{\psi}_a X_{a b} \psi_b \rangle \;,
\end{equation}
where
\begin{equation}\label{stataverage}
  \langle \cdots \rangle \equiv \frac{1}{\cal Z} \int D(\overline{\psi} \psi) \; ( \cdots ) \; e^{- S[\overline{\psi} , \psi ; F]} \;,
\end{equation}
with the partition function ${\cal Z} = \int D(\overline{\psi} \psi) \; \exp\{- S[\overline{\psi} , \psi ; F]\}$, and a Wick rotation was performed, $t \to i \tau$.

In our case, the Euclidean action $S [\overline{\psi}, \psi; A, \phi]$, is obtained from (\ref{action samples}), hence the response functions are\footnote{Notice that on the second equality we are omitting terms proportional to $\langle j^{\mbox{em}}_{\mu} \rangle =\frac{1}{\cal Z}\frac{\delta}{\delta A_{\mu}}\Big{|}_{A=0} \cal{Z}$, as well as $\langle j^{\mbox{tor}} \rangle =\frac{1}{\cal{Z}}\frac{\delta}{\delta \phi}\Big{|}_{\phi=0}\cal{Z}$. On this, see \cite[p.~370-371]{Altland-Simons} and Section \ref{section_nonlinear}.}
\begin{subequations}
\begin{align}
  \chi^{\mbox{em}}_{\mu \nu} (x,x') = \frac{\delta^2 }{\delta A_\mu (x) \delta A_\nu (x')} \Bigg{|}_{A=0} \ln{\cal Z} &= \frac{1}{\cal Z} \frac{\delta^2}{\delta A_\mu (x) \delta A_\nu (x')} \Bigg{|}_{A=0}  {\cal Z}  \sim \langle \hat{j}^{\mbox{em}}_\mu (x) \hat{j}^{\mbox{em}}_{\nu} (x') \rangle \;, \\
  \chi^{\mbox{tor}} (x,x') =  \frac{\delta^2 }{\delta \phi (x) \delta \phi (x')}  \Bigg{|}_{\phi=0} \ln{\cal Z} &=  \frac{1}{\cal Z} \frac{\delta^2 }{\delta \phi (x) \delta \phi (x')}  \Bigg{|}_{\phi=0} {\cal Z}
  \sim \langle \hat{j}^{\mbox{tor}} (x) \hat{j}^{\mbox{tor}} (x') \rangle \;, \\
  \chi^{\mbox{torem}}_{\mu} (x,x') =  \frac{\delta^2 }{\delta A_\mu (x) \delta \phi (x') }  \Bigg{|}_{A=\phi=0} \ln{\cal Z} &= \frac{1}{\cal Z} \frac{\delta^2}{\delta A_\mu (x) \delta \phi (x')}  \Bigg{|}_{A=\phi=0} {\cal Z}
  \sim \langle \hat{j}^{\mbox{em}}_\mu (x) \hat{j}^{\mbox{tor}} (x') \rangle \;,
\end{align}
\end{subequations}
which give, respectively, the macroscopic quantities
\begin{subequations}
\begin{align}
  j^{\mbox{em}}_{\mu} (x) &=  \int d^3 x' \chi^{\mbox{em}}_{\mu \nu} (x,x') A^\nu (x') \;, \\
  j^{\mbox{tor}} (x)    &=  \int d^3 x' \chi^{\mbox{tor}} (x,x') \phi (x') \;, \\
  j^{\mbox{torem}}_\mu (x)  &=  \int d^3 x' \chi^{\mbox{torem}}_{\mu} (x,x') \phi (x') \;.
\end{align}
\end{subequations}

With these, we can now move the first steps toward rephrasing the one-particle/classical picture presented above in terms of macroscopic many-body quantities. The former entail the microscopic elements of the (high energy/emergent analog) model, while macroscopic many-body quantities are ready for being used to design suitable experiments.

Let us now combine the two response functions to $A_\mu$ and to $\phi$, to have one vector and one scalar response
\begin{eqnarray}
  j_\mu (x)  &\equiv&  \int d^3 x' \left[ \chi^{\mbox{em}}_{\mu \nu} (x,x') A^\nu (x') + \chi^{\mbox{torem}}_{\mu} (x,x') \phi (x') \right] \;, \label{jcomb1}\\
  j (x)      &\equiv&  \int d^3 x' \left[ \chi^{\mbox{torem}}_{\mu} (x,x') A^\mu (x')  + \chi^{\mbox{tor}} (x,x') \phi (x') \right] \label{jcomb2} \;,
\end{eqnarray}
where, as required for gauge invariance and current conservation \cite[page~390]{Altland-Simons}, for an arbitrary function $\alpha(x)$, it is demanded that
\begin{equation*}
 \chi^{\mbox{em}}_{\mu \nu}(x,x')  \, \partial^{\nu}\alpha(x')=\chi^{\mbox{torem}}_{\nu} (x,x') \, \partial^{\nu}\alpha(x')=0 \;.
\end{equation*}
The actual realization of the time-loop (must come from the ``em'' part) with a torsion-induced holonomy (must come from the ``tor'' part), requires specific settings that we do not provide in this qualitative analysis.

The first of such settings is that, if no electromagnetic field is around to excite the particle-hole pair necessary for the t-loop, then the pure torsional contribution must be null
\begin{equation}\label{request1}
  \int d^3 x' \chi^{\mbox{tor}} (x,x') \phi (x') \to 0 \;.
\end{equation}
To prove the above, one will need details of actual structure of $\phi (x) = \epsilon_{\mu \nu \lambda} T^{\mu \nu \lambda}$, i.e., as made out of torsion, whereas we are dealing with it here merely as a scalar quantity.

On the other hand, the effects we are looking for must be encoded into the \textit{mixed response function} $\chi^{\mbox{torem}}_{\mu} (x,x')$. To have an idea of what these terms might mean, let us briefly recall the meaning of the well-known electromagnetic response function, for time and space translational invariant systems, for which
\begin{equation}\label{emlinresponsegen}
  j_\mu (\vec{p}, \omega) = \chi^{\mbox{em}}_{\mu \nu} (\vec{p}, \omega) A^{\nu} (\vec{p}, \omega) \,.
\end{equation}
Focusing on the spatial components (actual electric current) we have, essentially, two cases: the longitudinal response
\begin{equation}\label{emlinresponselong}
  \vec{j}_i (\vec{p}, \omega) = \vec{\chi}^{\mbox{em}}_{i 0} (\vec{p}, \omega) V (\vec{p}, \omega) + \chi^{\mbox{em}}_{i i} (\vec{p}, \omega) \vec{A}_i (\vec{p}, \omega) \,,
\end{equation}
(the $i$ index in the last term is not summed), and the transverse response
\begin{equation}\label{emlinresponsetrans}
  \vec{j}_i (\vec{p}, \omega) = \chi^{\mbox{em}}_{i j} (\vec{p}, \omega) \vec{A}_{j} (\vec{p}, \omega) \,,
\end{equation}
with $i \neq j$.

\subsection{Longitudinal response. External electric field}
\label{section_observable E}

The longitudinal response is what we should expect when a weak electric field is applied to the material. From
$\vec{E} (\vec{x}, t)= \vec{\nabla} V (\vec{x}, t) + \partial_t \vec{A} (\vec{x}, t)$, we get $\vec{E} (\vec{p}, \omega) = \vec{p} \, V (\vec{p}, \omega) +
\omega \vec{A} (\vec{p}, \omega)$. By setting $\chi^{\mbox{em}}_{i i} (\vec{p}, \omega) = \sigma (\vec{p}, \omega) \, \omega$, and $\chi^{\mbox{em}}_{o i} (\vec{p}, \omega) = \sigma (\vec{p}, \omega) \, \vec{p}_i$, we have the Ohm's law
\begin{equation}\label{Ohm}
  \vec{j} (\vec{p}, \omega)
  = \sigma (\vec{p}, \omega) \, (\vec{p} V (\vec{p}, \omega) + \omega \vec{A} (\vec{p}, \omega) )
  = \sigma (\vec{p}, \omega) \, \vec{E} (\vec{p}, \omega) \;.
\end{equation}

Therefore, along the same lines of what just recalled for the electromagnetic linear response, the Occam's razor would suggest the \textit{Ansatz}
\begin{equation}\label{linearTorsion}
\chi^{\mbox{torem}}_{o i} (\vec{p}, \omega) \equiv \tau (\vec{p}, \omega) \, \vec{p}_i \;,
\end{equation}
for the linear torsion response. Here, the conductivity $\tau$ that, in general, differs from the electric conductivity $\sigma$, and the explicit dependence from $\vec{p}_i$ is there due to indices structure of the response function (just like it happens for $\chi^{\mbox{em}}_{o i}$ above).

This produces a departure from the Ohm's law above, obtained from (\ref{jcomb1})
\begin{equation}
  \vec{j} \sim \sigma \, (\vec{p} \, V  + \omega \, \vec{A} ) + \vec{\chi}^{\mbox{torem}} \phi
  = \vec{p}  \, ( \sigma V  + \tau \phi ) + \omega \, \sigma \vec{A} \label{OhmTor} \;,
\end{equation}
written in Fourier space, $(\vec{p}, \omega)$.

\begin{figure}
\begin{center}
\includegraphics[width=0.5\textwidth,angle=0]{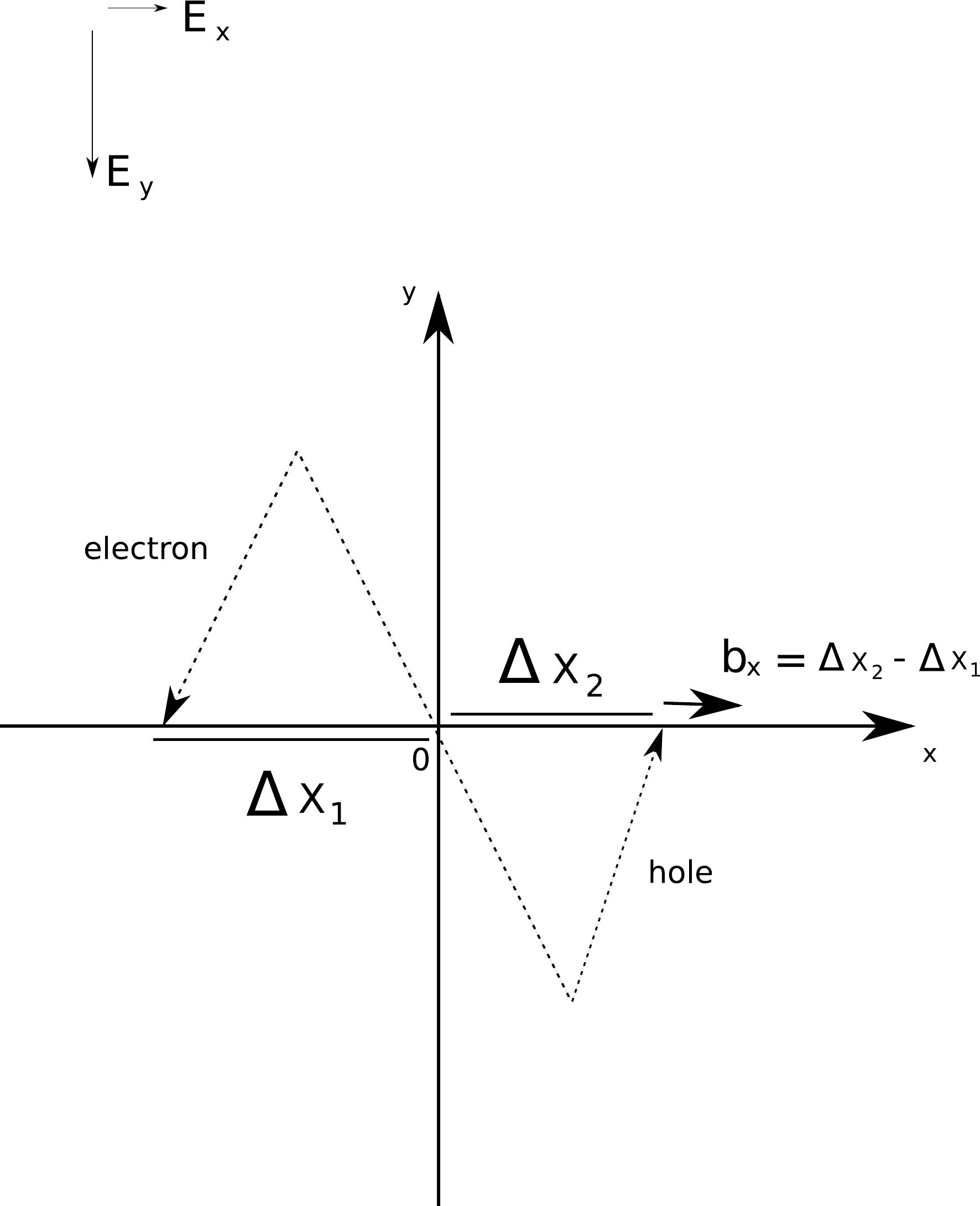}
\end{center}
\caption[3pt]{{An external electric field with components $E_{y}$ and $E_{x}$ is applied to the graphene sheet at $t=0$. The $E_{y}$ is flipped at $t=t^{*}$. We see the electron (hole) trajectory describes the upper-left (lower-right) triangle at $t=2t^{*}$. The total failure to come back to the same point is $\Delta x_{1}+\Delta x_{2}$. If $\Delta x_{1}\neq\Delta x_{2}$, then there could be a Burgers vector along the $x-$direction.}}
\label{x-y-electric}
\end{figure}

This is the many-body/macroscopic manifestation of the one-particle/classical picture of Fig. \ref{x-y-electric}. There the idea is to use an electric field along $y$ for some time $t^{*}$, during which the electron (hole) will move vertically along the positive (negative) $y$ axis. Then, at $t=t^*$, we abruptly flip the electric field $E_{y}$. Now the electron (hole) starts the journey to come back to the initial position $y=0$, that will eventually be reached at $t=2t^{*}$. However, in order to make it easier to spot the effect of a nonzero $b_x$, one should also add a small $x$ component $E_x$, there for all times $t$. Due to that, the electron (hole) also moves to the left (right). At time $t=2t^{*}$, the electron (hole) will have a horizontal separation of $\Delta x_{1}$ ($\Delta x_{2}$). All in all, if the difference $b_{x}=\Delta x_{1}-\Delta x_{2}$ is nonzero, this could point out the existence of a Burgers vector along the $x-$ direction. As in the $t-y$ plane, the particle performed a loop, and in the $x-y$, the particle-antiparticle pair does not come back to the same point. All this idea is depicted in Fig.\ref{x-y-electric}.

A flaw in this procedure is that, even when there is no defect, the particle-antiparticle pair never comes back to the same point. Indeed, after one loop in the $t-y$- plane, there is a difference of $\Delta x_{1}+\Delta x_{2}$, showing the failure to close the loop, regardless of whether the Burgers vector is zero or not. In other words, the so called \textit{contrast} (the ratio between presence and absence of signal) would be close to $0$, on the contrary to the ideal case, where a large contrast would be necessary to make observable the effect we are looking for.

\subsection{Transverse response. External magnetic field}
\label{section_observable B}

\begin{figure}
\begin{center}
\includegraphics[width=0.7\textwidth,angle=90]{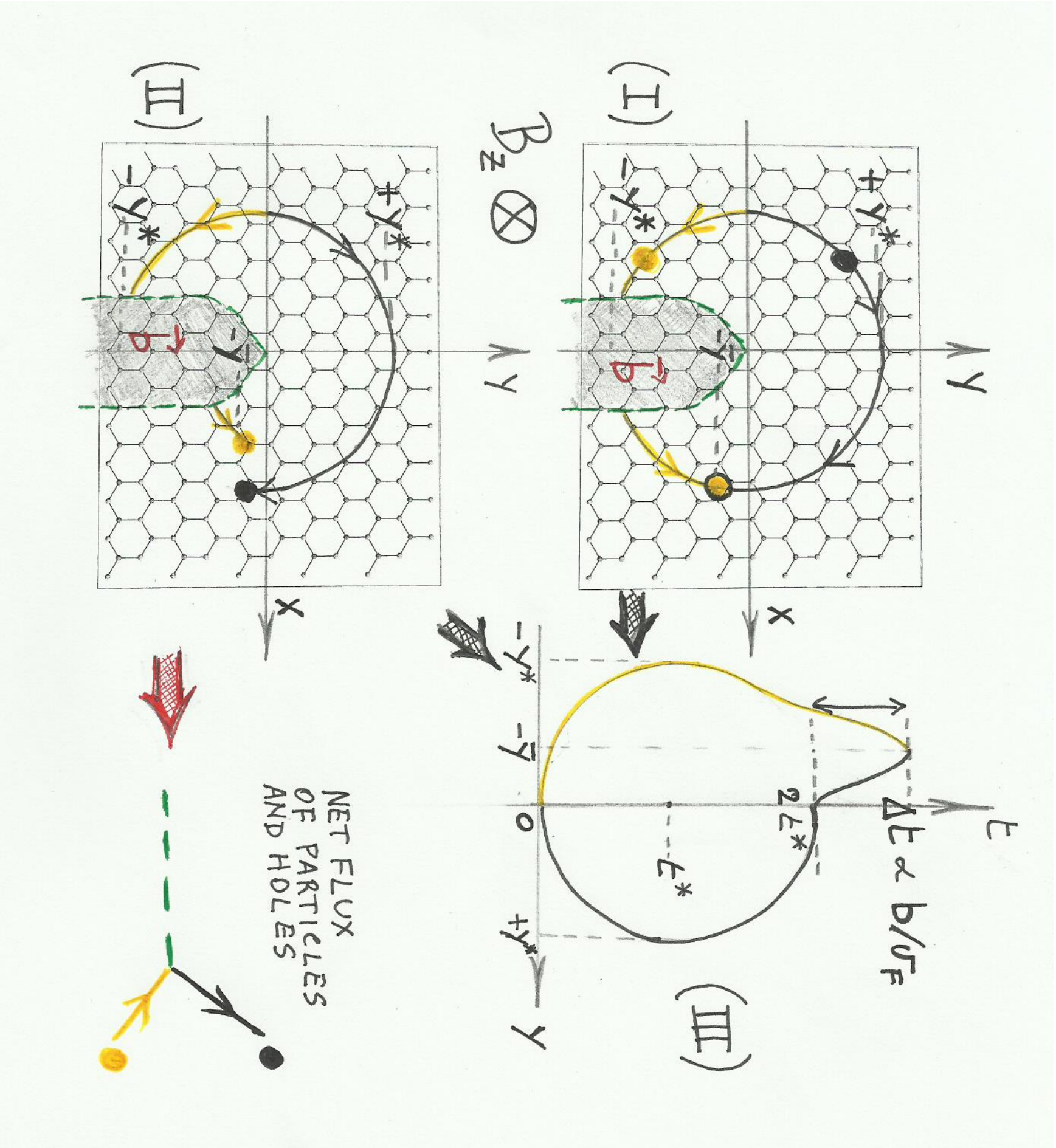}
\end{center}
\caption{Torsion/dislocation-induced idealized deformations of the idealized \textit{time-loop}. On the left, two possible effects of a magnetic field pointing into the plane $(x,y)$, in the presence of some nonzero dislocations, indicated with the shadowed region. Both in (I) and in (II), the antiparticle/hole travels through the shadowed region, that, although not necessarily so, can be thought of as buckling out of the plane, and deformed. The disturbance delays when the $y$-coordinate of particle and antiparticle is again the same ($-\bar{y}$ here). Therefore, both (I) and (II) produce the deformed \textit{time-loop} in the $(t,y)$-plane of (III). Nonetheless, it is only when particle and antiparticle do not meet, see (II), by a mismatch of their $x$-coordinate after a turn (related to the Burgers vector) that this produces a current, whose field theoretical description is represented in the depicted Feynman graph.}%
\label{Fig4NETFLUXandVERTEX}%
\end{figure}

On the other hand, the transverse electromagnetic current one obtains from $\chi^{\mbox{em}}_{ij}$, is nothing else than the \textit{Hall current}, that is the response to a vector potential $\vec{A} = B \vec{x} \times \vec{z}/z$, generating a magnetic field perpendicular to the plane of the Dirac material. In this case, one sees that the response function, containing the microscopic elements of the theory, again gives a measurable quantity that is the Hall conductivity $\sigma_{x y} = - \sigma_{y x}$, with
\begin{equation}\label{hallconductivity}
  \sigma_{i j} (x) \sim \frac{1}{\omega} \int d^2 x \, d^2 x' \, \chi^{\mbox{em}}_{i j} (x, x').
\end{equation}
With reference to equation (\ref{jcomb1}), we then see that the ideal realization of what represented in Fig.\ref{Fig4NETFLUXandVERTEX}, is to apply a magnetic field, that will separate positive and negative charge carriers (as customary in the Hall effect). The one-particle picture is that such a field, when of sufficient strength, excites a pair particle-hole out of the vacuum, and both particle and hole turn around the dislocation line, in the $(x,y)$-plane, as shown in Fig.\ref{Fig3TIMELOOP}. The corresponding \textit{time-loop} in the $(y,t)$-plane (supposing that the Burgers vector is directed along $x$, like in Fig.\ref{Fig1}), is necessarily deformed, the deformation being proportional to the magnitude of the Burgers vector, $\Delta t \propto b / v_F$.

In Fig.\ref{Fig4NETFLUXandVERTEX}, we depict two possibilities, (I) and (II), both giving the deformed \textit{time-loop} in the $(y,t)$-plane (III), but only (II) truly includes the required holonomy, that should give rise to a net flux of particles and antiparticles, giving meaning to the vertex $\overline{\psi}\gamma^{5}\phi \psi$, hence directly related to the dislocations present in the material.

In the language of the response function, with reference to Eq.~(\ref{jcomb1}), we should have two contributions to the combined current. One contribution is entirely electromagnetic, and is a transverse current directed along $y$. The other contribution, entirely due to the response to torsion, could be engineered to be along the $x$-direction. In summary
\begin{equation}\label{currentB}
  \vec{j}^{\mbox{em}} = (0, j_y^{\mbox{em}}) \quad {\rm and } \quad   \vec{j}^{\mbox{torem}} = (j_{x}^{\mbox{torem}}, 0) \,.
\end{equation}
This appears to be the most promising way to spot the effect, for at least two reasons. First, as well known, the Hall current is very sensitive to the different types of carriers, (quasi) electrons and (quasi) holes. This is crucial for our time-loop. Second, the contrast here is, in principle, infinite because along the $x$ direction the only contribution would be the one induced by torsion, $j_{x}^{\mbox{torem}}\neq 0$ vs $j_{x}^{\mbox{em}} = 0$, therefore even the smallest effect of a nonzero $j_{x}^{\mbox{torem}}$ should be visible (of course, within limits imposed by instruments, noise, etc.).

\subsection{Yet one more possibility}

One more way to spot the effect, in principle, would be to refer to the $\mu = 0 = \nu$ contribution of equation (\ref{jcomb1}), and compare to the $\mu = 0$ component of (\ref{jcomb2}). In this case we should see an impact of the torsion on the response density, or, in other words, on the way the charges distribute. Again, though, the contrast would not be ideal.

\section{Necessity for non-linear response}
\label{section_nonlinear}

The simple, semi-qualitative analysis presented in the previous Section, is only indicative, and cannot be pushed too far. In fact, in our model, described microscopically by the action \eqref{action samples}, we can indeed produce a prediction based on the charge conjugation invariance of that emergent relativistic theory. Such prediction is that
\begin{equation}\label{firstFurry}
    \chi^{\mbox{torem}}_{\mu} (x,x') \sim \langle \hat{j}^{\mbox{em}}_{\mu} (x) \hat{j}^{\mbox{tor}} (x') \rangle \equiv 0 \;.
\end{equation}
This is nothing more than an instance of the Furry's theorem of quantum field theory \cite{Peskin}, that in QED reads\footnote{While $C^{-1} \hat{j}^{\mbox{tor}} C = \hat{j}^{\mbox{tor}}$, hence we can have any number of them in the vacuum expectation value (VEV), for the other current $C^{-1} \hat{j}^{\mbox{em}} C = - \hat{j}^{\mbox{em}}$, so that $\langle \Omega | \hat{j}^{\mbox{em}}_{\mu} | \Omega \rangle =
\langle \Omega | C^{-1} \hat{j}^{\mbox{em}}_{\mu} C | \Omega \rangle = - \langle \Omega | \hat{j}^{\mbox{em}}_{\mu} | \Omega \rangle \equiv 0$, where we changed notation for the VEV, and we used charge conjugation invariance of the vacuum $|\Omega\rangle = C |\Omega\rangle$.}
\begin{equation}\label{FurryEMgeneral}
    \chi^{\mbox{em}}_{\mu_{1} ... \mu_{2n+1}} (x_{1}, ..., x_{2n+1}) \sim \langle \hat{j}^{\mbox{em}}_{\mu_{1}} (x_{1}) \cdots \hat{j}^{\mbox{em}}_{\mu_{2n+1}} (x_{2n+1}) \rangle = 0 \;,
\end{equation}
and for us implies
\begin{equation}\label{Furry_ours}
    \chi^{\mbox{torem}}_{\mu_{1} ... \mu_{2n+1}} (x_{1}, ..., x_{2n+1}, y_{1}, ..., y_{m}) \sim \langle \hat{j}^{\mbox{em}}_{\mu_{1}} (x_{1}) \cdots \hat{j}^{\mbox{em}}_{\mu_{2n+1}} (x_{2n+1})
    \hat{j}^{\mbox{tor}} (y_1) \cdots \hat{j}^{\mbox{tor}} (y_{m})\rangle = 0 \;.
\end{equation}
This result does not mean that we have to find a completely different approach, or that the effects we are looking for cannot be seen in this language. This result simply means that we need to move to the nonlinear response regime.

Indeed, our general expectations could be formalized as two kinds of requests on the functional expansion of the response (\ref{X[F]}): (a) terms with an odd number of derivatives $\delta/\delta A_{\mu}$, and (b) terms with only $\delta^n /\delta \phi(x_1) \cdots \delta \phi(x_n)$ must not be there. The two conditions are based on different criteria. The first one is strictly related to the validity of the emergent analog relativistic model based on the action $S[\overline{\psi}, \psi, A, \phi]$. The second is a request that needs be obtained from the torsional nature of the field $\phi$, that we did not include in the previous analysis.

With these considerations, the first nonzero contribution would be
\begin{equation}
  j^{\mbox{torem}}_{\mu} (x) =  \int d^3 x' d^3 x'' \chi^{\mbox{torem}}_{\mu \nu} (x,x',x'') A^{\nu} (x') \phi (x'') \;.
\end{equation}
with
\begin{equation}\label{firstnonzeroresponse}
\chi^{\mbox{torem}}_{\mu \nu} (x,x',x'') =  \frac{1}{{\cal Z}} \frac{\delta^{3}}{\delta A_\mu (x) \delta A_\nu (x') \delta \phi (x'')}\Bigg{|}_{A=0} {\cal Z}
  \sim \langle \hat{j}^{\mbox{em}}_{\mu} (x) \hat{j}^{\mbox{em}}_{\mu} (x') \hat{j}^{\mbox{tor}} (x'') \rangle \;.
\end{equation}

To simplify the discussion, let us focus on the time dependance only, an on the current rather than the charge response
\begin{equation}
  j^{\mbox{torem}}_i (t)  =  \int dt' \, dt'' \, \chi^{\mbox{torem}}_{i j} (t,t',t'') \, A^{j} (t') \, \phi (t'') \;,
\end{equation}
that, in terms of Fourier components\footnote{Our Fourier transform convention for a function $f:\mathbb{R}^{n}\to\mathbb{R}$ is $f(x)=\int \frac{d^{n}k}{(2\pi)^{n/2}}\, e^{-ik\cdot x}\,f(k)$, and $f(k)=\int \frac{d^{n}x}{(2\pi)^{n/2}}\, e^{ik\cdot x}\,f(x)$.}, reads
\begin{equation}
  j^{\mbox{torem}}_i (\omega)  =   \int d\omega' d\omega'' \chi^{\mbox{torem}}_{i j} (\omega, - \omega',- \omega'') A^{j} (\omega') \phi (\omega'') \;.
\end{equation}

By regarding this as a process of a stimulated emission of frequency $\omega$, from inputs of frequencies $\omega'$ and $\omega''$, the conservation of energy implies $\omega = \omega' + \omega''$. If, furthermore, $\omega' = \omega''$, then $\omega = 2 \omega'$, i.e., the system responds to a probe of given frequency generating higher harmonics (second harmonic in this example).

Therefore, for the experiment we are looking for, we can resort to a well developed technique, the high-order harmonic generation (HHG), able to characterize structural changes both in atoms and molecules and, more recently, bulk materials (for a recent review see e.g.~\cite{StanislavRevModPhys}).  HHG is a nonlinear optical phenomena in which the frequency of the laser light that drives the system is converted into its integer multiples. Harmonics of very high orders are generated when atoms, molecules and, recently, solid materials, are exposed to intense (usually near-infrared) and short (within the femtosecond domain) laser pulses. Particularly, the spectrum from this process consists of a plateau, where the harmonic intensity is nearly constant over many orders, and it suddenly ends up, at the so-called HHG cutoff. HHG is considered nowadays as one of the best methods to both produce ultrashort coherent light covering a wavelength range from the vacuum ultraviolet to the soft x-ray region and to obtain atomic, molecular and condensed matter structural information with, unique, nanometer spatial resolution. The development of HHG has opened new research areas such as attosecond science and nonlinear optics in the extreme ultraviolet (XUV) region~\cite{KrauszIvanov}. In the last few years, the subject of HHG from solid-state samples has attracted considerable attention~\cite{GhimireNatPhy2011, VampaJPB2017, EOsikaPRX2017}. In particular, it is now experimentally possible to disentangle the intra-band and inter-band currents, and how to use HHG to characterize structural information such as the energy dispersions~\cite{VampaNat2015, VampaPRB2015, VampaPRL2015}. Very recently, Berry-phase effects have been explored in topologically-trivial materials, through experimental studies of HHG in atomically-thin semiconductors~\cite{Liu2017} and in quasi-2D models~\cite{Luu2018}, where the sensitivity of harmonic emission to symmetry breaking (specifically, the breaking of inversion symmetry in monolayer MoS$_2$ and $\alpha$-quartz) is shown via the presence of even harmonics.

In our scheme, the intra-band harmonics, governed by the intra-band (electron-hole) current, will be strongly modified, depending on the presence, or not, of dislocations.
Indeed, one immediate impact of the previous discussion on the structure of the nonlinear response, would be that, the torsion-induced holonomy of the time-loop could manifest itself through specific patterns of suppression and generation of higher harmonics.

\section{Conclusions}
\label{section_conclusions}

We conclude that, when time is duly included in the emergent relativistic-like picture of Dirac materials, the geometric obstruction to describe the effects of dislocations in terms of a suitable coupling with torsion, within the (2+1)-dimensional field theoretical description of the $\pi$-electrons dynamics, can be overcome. This is not a proof that torsion indeed describes dislocations in these cases, and surely problems remain to be addressed, like a unique assignment of torsion to a given distribution of Burgers vectors. Nonetheless, when this is possible, our suggestion here opens the doors to the use of these materials as analogs of many important theoretical scenarios where torsion plays a role.

Although our paper is theoretical, we moved the first steps toward testing the realization of these scenarios. We have envisaged some kinds of \textit{Gedankenexperiments} on \textit{time-loop} that could spot the presence of edge dislocations, routinely produced in Dirac materials. The effect must be based on the interplay between an external electromagnetic field (necessary to excite the pair particle-hole that realizes the time-loops), and a suitable distribution of dislocations described as torsion (that will be responsible for a measurable holonomy in the time-loop).

Our general analysis here establishes that we need to move to a nonlinear response regime. In particular, we speculate that in a HHG technique, the structure of such response, would include manifestation of the torsion-induced holonomy of the time-loop through specific patterns of suppression and generation of higher harmonics. This sounds promising, for an experimental finding, as the laser-graphene interaction, controlling electron dynamics on an unprecedented precision scale, is the focus of intense studies, both theoretical and experimental, see, e.g., \cite{PhysRevLett.121.207401, natureLaserGraphene}. Nonetheless, our results here need further detailed analysis, that is beyond the scope of this paper, and we intend to perform in future work.

%-------------------------------------------------------------------------------------------------------------------%-------------------------------------------------------------------------------------------------------------------
%Acknowledgments
%-------------------------------------------------------------------------------------------------------------------
\section*{Acknowledgments}

We thank Tadzio Levato for many informative discussions on the physics of lasers, and more. M.~F.~C. acknowledges the project Advanced research using high intensity laser produced photons and particles (CZ.02.1.01/0.0/0.0/16\_019/0000789) from European Regional Development Fund (ADONIS), the Spanish Ministry MINECO (National Plan
15 Grant: FISICATEAMO No. FIS2016-79508-P, SEVERO OCHOA No. SEV-2015-0522, FPI), European Social Fund, Fundaci\'o Cellex, Generalitat de Catalunya (AGAUR Grant No. 2017 SGR 1341 and CERCA/Program), ERC AdG NOQIA, and the National Science Centre, Poland-Symfonia Grant No. 2016/20/W/ST4/00314.
. P.~P. is supported by the project High Field Initiative (CZ.02.1.01/0.0/0.0/15\_003/0000449) from the European Regional Development Fund. A.~Z. acknowledges support from the ``Albert Einstein Center for Gravitation and Astrophysics" and the COST Action MP1405.

%-------------------------------------------------------------------------------------------------------------------%-------------------------------------------------------------------------------------------------------------------
% Appendices
%-------------------------------------------------------------------------------------------------------------------

\appendix

\section{Minimal spinor coupling with torsion}
\label{appendix_torsion}

Here we will recall the well-known argument, according to which, spinors are only coupled with the \textit{totally antisymmetric} part of the torsion, in the minimal coupled prescription \cite{Shapiro}.

Suppose we have the following Hermitian and local Lorentz invariant action (here we used natural units $[\hbar]=[c]=1$)
\begin{eqnarray}\label{Dirac_action}
S&=&\frac{i}{2} \int d^{3}x \sqrt{|g|} \, \left(\overline{\Psi}\gamma^{\mu}\overrightarrow{D}_{\mu}\Psi-\overline{\Psi}\overleftarrow{D}_{\mu}\gamma^{\mu}\Psi\right) \\
&=&\frac{i}{2} \int d^{3}x \sqrt{|g|} \, \left(\overline{\Psi}\gamma^{\mu}\overrightarrow{D}_{\mu}\Psi-\partial_{\mu}\overline{\Psi}\gamma^{\mu}\Psi+\frac{i}{2}\omega^{ab}_{\mu} \overline{\Psi}\mathbb{J}_{ab}\gamma^{c}\Psi E_{c}^{\mu}\right) \;, \nonumber
\end{eqnarray}
where the covariant derivatives\footnote{Here we work in the two-index notation for the Lorentz generators and the spin connection to keep the discussion as general as possible. Of course we can comeback to the dual one-index notation in three dimensions.}
\begin{eqnarray*}
\overrightarrow{D}_{\mu}\Psi&=&\partial_{\mu}\Psi+\frac{i}{2}\omega^{ab}_{\mu}\mathbb{J}_{ab}\Psi\;,\\
\overline{\Psi}\overleftarrow{D}_{\mu}&=&\partial_{\mu}\overline{\Psi}-\frac{i}{2}\overline{\Psi}\mathbb{J}_{ab}\omega^{ab}_{\mu}\;,
\end{eqnarray*}
contain the contorsion part inside the spin connection, i.e., $\omega^{ab}=\mathring{\omega}^{ab}+\kappa^{ab}$. With this convention, we have $de^{a}-\tensor{\mathring{\omega}}{^{a}_{b}}e^{b}=0$ and $T^{a}=-\tensor{\kappa}{^{a}_{b}}e^{b}$. To relate the last expression to the torsion tensor, $T^\lambda_{\mu \nu} = \Gamma^\lambda_{\mu \nu} - \Gamma^\lambda_{\nu \mu}$, one uses $\tensor{T}{_{\mu}^{\lambda} _{\nu}} =E^\lambda_a \tensor{\kappa}{^{a}_{\nu}_{b}} e^b_\mu - E^\lambda_a \tensor{\kappa}{^{a}_{\mu}_{b}} e^b_\nu$.

In order to obtain the field equations for $\Psi$, we should vary the action under $\overline{\Psi}$. Therefore, we must integrate by parts the second term of \eqref{Dirac_action}.
\begin{eqnarray}\label{action_split}
S&=&\frac{i}{2} \int d^{3}x \, \sqrt{|g|} \, \left(\overline{\Psi}\gamma^{\mu}{D}_{\mu}\Psi+\overline{\Psi}E_{a}^{\mu}\gamma^{a}\partial_{\mu}\Psi+\frac{i}{2}\omega^{bc}_{\mu} \overline{\Psi}\gamma^{a}\mathbb{J}_{bc}\Psi E_{a}^{\mu}+\frac{i}{2}\omega^{bc}_{\mu} \overline{\Psi}[\mathbb{J}_{bc},\gamma^{a}]\Psi E_{a}^{\mu}\right) \nonumber \\
&&+ \frac{i}{2} \int d^{3}x \partial_{\mu}\left(\sqrt{|g|}E_{a}^{\mu}\right)\overline{\Psi}\gamma^{a}\Psi  \; + \; \mbox{BT} \; \nonumber \\
&=&\frac{i}{2} \int d^{3}x \, \sqrt{|g|} \, \left(2\overline{\Psi}\gamma^{\mu}{D}_{\mu}\Psi+\frac{i}{2}\omega^{bc}_{\mu} \overline{\Psi}[\mathbb{J}_{bc},\gamma^{a}]\Psi E_{a}^{\mu}\right) \nonumber \\
&&+ \frac{i}{2} \int d^{3}x \partial_{\mu}\left(\sqrt{|g|}E_{a}^{\mu}\right)\overline{\Psi}\gamma^{a}\Psi + BT    \; ,
\end{eqnarray}
where $D_{\mu} \equiv \overrightarrow{D}_{\mu}$, and $\mbox{BT}$ is a boundary term, which could have some role in defining conserved charges, but we shall not take it into account here. Let us manipulate the last term in the first integral in \eqref{action_split},
\begin{equation}\label{relation}
\frac{i}{2}\omega^{bc}_{\mu}[\mathbb{J}_{bc},\gamma^{a}] =-\frac{1}{2}\left(\omega^{ba}_{\mu}\gamma_{b}-\omega^{ac}_{\mu}\gamma_{c}\right)=\tensor{\omega}{^{a}_{\mu}_{b}}\gamma^{b}\;,
\end{equation}
where in the first equality we used the property $\left[\gamma^{a},\mathbb{J}_{bc}\right]= i \left(\gamma_{c}\delta^{a}_{b}-\gamma_{b}\delta^{a}_{c}\right)$.

Now,
\begin{equation*}
\frac{i}{2}\omega^{bc}_{\mu}\overline{\Psi}[\mathbb{J}_{bc},\gamma^{a}]\Psi E_{a}^{\mu}=\tensor{\omega}{^{a}_{\mu}_{b}}E_{a}^{\mu}\overline{\Psi}\gamma^{b}\Psi=\tensor{\omega}{^{a}_{\mu}_{b}}E_{a}^{\mu}e^{b}_{\nu}\overline{\Psi}\gamma^{\nu}\Psi=E_{a}^{\mu}E_{b}^{\nu}\tensor{\omega}{^{a}_{\mu}_{c}}e^{c}_{\nu}\overline{\Psi}\gamma^{b}\Psi \;.
\end{equation*}
We observe here that the term
\begin{equation}\label{zero_term}
E_{a}^{\mu}E_{b}^{\nu}\tensor{\omega}{^{a}_{\nu}_{c}}e^{c}_{\mu}\overline{\Psi}\gamma^{b}\Psi=\delta_{a}^{c}E_{b}^{\nu}\tensor{\omega}{^{a}_{\nu}_{c}}\overline{\Psi}\gamma^{b}\Psi=0\;,
\end{equation}
where in the last equality we used the antisymmetry of $\omega^{ab}$. Therefore, we can add safely the term \eqref{zero_term} to the action. So far, we have
\begin{equation*}
S=\frac{i}{2} \int d^{3}x \sqrt{|g|} \, \left(2\overline{\Psi}\gamma^{\mu} D_{\mu}\Psi - E_{a}^{\mu}E_{b}^{\nu}\left(\tensor{\omega}{^{b}_{\mu c}}e^{c}_{\nu}-\tensor{\omega}{^{b}_{\nu c}}e^{c}_{\mu}\right)\overline{\Psi}\gamma^{a}\Psi\right)+ \frac{i}{2} \int d^{3}x \partial_{\mu}\left(\sqrt{|g|}E_{a}^{\mu}\right)\overline{\Psi}\gamma^{a}\Psi \;.
\end{equation*}
Now, we move to the second integral in \eqref{action_split}. First of all, remember that \cite{Nakahara} $\sqrt{|g|}=|e|$, where for $|e|$ we understand the determinant of the dreibein, i.e., $|e|\epsilon_{\mu\nu\rho}=\epsilon_{abc}e^{a}_{\mu}e^{b}_{\nu}e^{c}_{\rho}$. So,
\begin{equation*}
\partial_{\mu}(\sqrt{|g|})=\partial_{\mu}|e|=\frac{1}{3!}\partial_{\mu}\left(\epsilon^{\mu\nu\rho}\epsilon_{abc}e^{a}_{\mu}e^{b}_{\nu}e^{c}_{\rho}\right)=\frac{1}{2}\epsilon^{\nu\rho\tau}\epsilon_{abc}\partial_{\mu}e^{a}_{\nu}e^{b}_{\rho}e^{c}_{\tau}\;.
\end{equation*}
Observe that the dreibein determinant fulfils the relation $\epsilon_{abc}e^{a}_{\mu}e^{b}_{\nu}=|e|\epsilon_{\mu\nu\rho}E_{c}^{\rho}$. Then,
\begin{equation*}
\partial_{\mu}(\sqrt{|g|})=\frac{|e|}{2}E_{a}^{\sigma}\partial_{\mu}e^{a}_{\nu}\epsilon^{\nu\rho\tau}\epsilon_{\rho\tau\sigma}=|e|E_{a}^{\nu}\partial_{\mu}e^{a}_{\nu}\;.
\end{equation*}
It is important the property,
\begin{equation*}
\partial_{\mu}(E_{a}^{\nu}e^{b}_{\nu})=0=e^{b}_{\nu}\partial_{\mu}E_{a}^{\nu}+E_{a}^{\nu}\partial_{\mu}e^{b}_{\nu} \Rightarrow  e^{b}_{\nu}\partial_{\mu}E_{a}^{\nu}=-E_{a}^{\nu}\partial_{\mu}e^{b}_{\nu} \Rightarrow \partial_{\mu}E_{a}^{\rho}=-E_{b}^{\rho}E_{a}^{\nu}\partial_{\mu}e^{b}_{\nu} \;,
\end{equation*}
or
\begin{equation*}
\partial_{\mu}E_{a}^{\mu}=-E_{a}^{\mu}E_{b}^{\nu}\partial_{\nu}e^{b}_{\mu}\;.
\end{equation*}
Finally, we can compute the second integrand in \eqref{action_split}, as
\begin{eqnarray*}
\partial_{\mu}\left(\sqrt{|g|}E_{a}^{\mu}\right)&=&E_{a}^{\mu}\partial_{\mu}(\sqrt{|g|})+\sqrt{|g|}\partial_{\mu}E_{a}^{\mu} \\
&=&\sqrt{|g|}E_{a}^{\mu}E_{b}^{\nu}\left(\partial_{\mu}e^{b}_{\nu}-\partial_{\nu}e^{b}_{\mu}\right)\;.
\end{eqnarray*}
The action can be regrouped as
\begin{eqnarray*}
S&=&\frac{i}{2} \int d^{3}x \sqrt{|g|} \, \left(2\overline{\Psi}\gamma^{\mu}{D}_{\mu}\Psi + E_{a}^{\mu}E_{b}^{\nu}\left(\partial_{\mu}e^{b}_{\nu}-\partial_{\nu}e^{b}_{\mu}-\tensor{\omega}{^{b}_{\mu d}}e^{d}_{\nu}+\tensor{\omega}{^{b}_{\nu d}}e^{d}_{\mu}\right)\overline{\Psi}\gamma^{a}\Psi\right) \\
&=&\frac{i}{2} \int d^{3}x \sqrt{|g|} \, \left(2\overline{\Psi}\gamma^{\mu}{D}_{\mu}\Psi + E_{a}^{\mu}E_{b}^{\nu}\tensor{T}{_{\mu}^{b}_{\nu}}\overline{\Psi}\gamma^{a}\Psi\right) \\
&=&i\int d^{3}x \sqrt{|g|} \, \left(\overline{\Psi}\gamma^{\mu}{D}_{\mu}\Psi + \frac{1}{2}\tensor{T}{_{\mu}^{\nu}_{\nu}}\overline{\Psi}\gamma^{\mu}\Psi\right) \;,
\end{eqnarray*}
which is the result given in equation  (2.33) of \cite{Shapiro}, but now adapted to three dimensions and our metric sign conventions. The last action is expressed in terms of the total covariant derivative $D_{\mu}$. If we separate the contorsion component from this covariant derivative, we have $\overline{\Psi}\gamma^{\mu} D_{\mu}\Psi = \overline{\Psi}\gamma^{\mu} \mathring{D}_{\mu}\Psi + \frac{i}{2} \overline{\Psi} \gamma^{\mu} \kappa_{\mu}^{a b} \mathbb{J}_{ab} \Psi$, where $\mathring{D}_{\mu}$ is the covariant derivative containing only the torsionless part of the connection. Let us focus on the term containing $\kappa^a_b$. First, we notice that
$\gamma^{\mu} \kappa_{\mu}^{a b} \mathbb{J}_{ab} = E_c^\mu \kappa_{\mu}^{a b} \gamma^c \mathbb{J}_{ab}$. Then, we use $\mathbb{J}_{ab} = \frac{i}{4} [\gamma_a, \gamma_b]$, and
\begin{equation*}
\gamma^c \mathbb{J}_{ab} = \frac{i}{2} \delta^c_a \gamma_b - \frac{i}{2} \delta^c_b \gamma_a
+ \frac{i}{2} \tensor{\epsilon}{^{c}_{a b}} \gamma^{0} \gamma^{1} \gamma^{2} \;.
\end{equation*}
With these
\begin{equation*}
\frac{i}{2} \overline{\Psi} \gamma^{\mu} \kappa_{\mu}^{a b} \mathbb{J}_{ab} \Psi =
- \frac{1}{2} \overline{\Psi} E_a^\mu \kappa_{\mu}^{a b} \gamma_b  \Psi
+ \frac{1}{4} \overline{\Psi} E_c^\mu \kappa_{\mu}^{a b} \tensor{\epsilon}{^{c}_{a b}} \gamma^{0} \gamma^{1} \gamma^{2}  \Psi \;.
\end{equation*}
If we now use, for instance, the (reducible) representation $\gamma^{\mu} = \left(
                    \begin{array}{cc}
                    \gamma^{\mu}_{+} & 0 \\
                    0 & \gamma^{\mu}_{-} \\
                    \end{array}
                    \right)                 $ of the Lorenz group \cite{IORIO2018265},
\begin{eqnarray}\label{gamma_matrices_twoDiracpoints}
\gamma^0 &=& \left(
                    \begin{array}{cc}
                    \sigma^3 & 0 \\
                    0 & \sigma^3 \\
                    \end{array}
                    \right)                   \; ,  \\
\gamma^{1}&=& \left(
                    \begin{array}{cc}
                    i \sigma^{2} & 0 \\
                    0 & - i \sigma^{2} \\
                    \end{array}
                    \right)                   \; ,  \\
\gamma^{2}&=& \left(
                    \begin{array}{cc}
                    - i \sigma^{1} & 0 \\
                    0 & - i \sigma^{1} \\
                    \end{array}
                    \right)                   \;,
\end{eqnarray}
we have a natural definition of $\gamma^{5}$ as
\begin{equation}\label{gamma5}
\gamma^{5} \equiv i \gamma^0 \gamma^1 \gamma^2 = \left(
                    \begin{array}{cc}
                    I_{2 \times 2} & 0 \\
                    0 & - I_{2 \times 2} \\
                    \end{array}
                    \right)                   \; .  \\
\end{equation}
Taking into account that $E_a^\mu \tensor{\kappa}{^{a}_{\mu}^{b}} \gamma_b = E_a^\mu \tensor{\kappa}{^{a}_{\mu}_{b}}  \gamma^b = E_a^\mu \tensor{\kappa}{^{a}_{\mu}_{b}} e_\rho^b \gamma^\rho =- T^\mu_{\mu \rho} \gamma^\rho =  T^\mu_{\rho \mu} \gamma^\rho$, and that, following similar steps $E_c^\mu {\kappa_{\mu}}^{a b} \tensor{\epsilon}{^{c}_{a b}}
= \frac{\epsilon^{\mu \nu \rho}}{|e|} e_\nu^a e_\rho^b {\kappa}_{a \mu b} = - \frac{\epsilon^{\mu \nu \rho}}{|e|}  T_{\mu \nu \rho}$, we obtain
\begin{equation*}
\frac{i}{2} \overline{\Psi} \gamma^{\mu} {\kappa_{\mu}}^{a b} \mathbb{J}_{ab} \Psi =
- \frac{1}{2} \overline{\Psi} T^\mu_{\rho \mu} \gamma^\rho \Psi
- \frac{i}{4} \frac{\epsilon^{\mu \nu \rho}}{|e|} T_{\mu \nu \rho} \overline{\Psi}\gamma^{5}\Psi \;.
\end{equation*}
Finally, we arrive to
\begin{eqnarray}\label{Dirac_action_totally_antisymmetric}
S & = & i\int d^{3}x |e| \left(\overline{\Psi}\gamma^{\mu}\mathring{D}_{\mu}\Psi
- \frac{1}{2} \overline{\Psi} T^\nu_{\mu \nu} \gamma^{\mu} \Psi
- \frac{i}{4} \frac{\epsilon^{\mu \nu \rho}}{|e|}\overline{\Psi} \gamma^{5} \Psi
+ \frac{1}{2}  \overline{\Psi} T^\nu_{\mu \nu} \gamma^{\mu} \Psi \right)  \nonumber \\
& = & i\int d^{3}x |e| \left(\overline{\Psi}\gamma^{\mu}\mathring{D}_{\mu}\Psi
- \frac{i}{4} \frac{\epsilon^{\mu \nu \rho}}{|e|} T_{\mu \nu \rho} \overline{\Psi}  \gamma^{5} \Psi \right) \\
& = & i\int d^{3}x |e| \left(\overline{\psi}_{+}\gamma^{\mu}_{+}\mathring{D}_{\mu}\psi_{+}+\overline{\psi}_{-}\gamma^{\mu}_{-}\mathring{D}_{\mu}\psi_{-}
- \frac{i}{4} \frac{\epsilon^{\mu \nu \rho}}{|e|} T_{\mu \nu \rho} \left(\overline{\psi}_{+} \psi_{+} - \overline{\psi}_{-} \psi_{-}\right)  \right) \;, \nonumber
\end{eqnarray}
with the spinor coupling only to the totally antisymmetric components of torsion.

Notice that the spinors associated to each Dirac point, $\psi_{+}$ and $\psi_{-}$, are decoupled even when the torsion is included. Therefore, the field equations, obtained by varying the action \eqref{Dirac_action_totally_antisymmetric} with respect to the independent fields $\overline{\psi}_{+}$ and $\overline{\psi}_{-}$ are
\begin{eqnarray}
\gamma^{\mu}_{+}\mathring{D}_{\mu}\psi_{+} - \frac{i}{4} \frac{\epsilon^{\mu \nu \rho} T_{\mu \nu \rho} }{|e|} \,  \psi_{+}  &=& 0 \;, \\
\gamma^{\mu}_{-}\mathring{D}_{\mu}\psi_{-} + \frac{i}{4} \frac{\epsilon^{\mu \nu \rho} T_{\mu \nu \rho} }{|e|} \,  \psi_{-}  &=& 0 \;,
\end{eqnarray}
respectively.

\section{Zero curvature and nonzero torsion}
\label{appendix_zero_curvature}

In this Appendix we use the notation of differential forms (practically, this means that there are no explicit Einstein indices $\mu, \nu, ...$). In the general case where we have torsion and curvature, the Lorentz spin-connection takes the form $\omega^{ab} = \mathring{\omega}^{ab} + \kappa^{ab}$. The first term contribution is the Riemannian or Levi-Civita connection, while the second one is the contortion. Correspondingly, the Lorentz curvature $R^{ab}=d\omega^{ab}-\tensor{\omega}{^{a}_{c}}\omega^{cb}$, can be split as
\begin{equation*}
R^{ab}=\mathring{R}^{ab} +  D\kappa^{ab} = \mathring{R}^{ab} + \mathring{D}\kappa^{ab} - \tensor{\kappa}{^{a}_{c}}\kappa^{cb} \;,
\end{equation*}
where $\mathring{R}^{ab}$ is the Riemannian curvature. In this work, we commit ourselves in a particular situation where the torsion contribution can be isolated from pure geometric curvature. Thus, we propose a situation where the Riemannian curvature is zero ($\mathring{R}^{ab}=0$), but $\kappa^{ab}\neq0$. This proposal is meaningful as $\kappa^{ab}$ transforms as a tensor under Lorentz transformations, therefore $\kappa^{ab}\neq0$ is independent of the selected frame. On the other hand, as $\mathring{R}^{ab}=0$, we can choose also a Lorentz frame where the torsionless spin connection is locally zero ($\mathring{\omega}^{ab}=0$).

%-------------------------------------------------------------------------------------------------------------------%-------------------------------------------------------------------------------------------------------------------
% References
%-------------------------------------------------------------------------------------------------------------------
\bibliographystyle{apsrev4-1}
\bibliography{libraryTorsion}
%\input{arXiv.bbl}
%-------------------------------------------------------------------------------------------------------------------
%-------------------------------------------------------------------------------------------------------------------
\end{document}